\documentstyle[12pt]{article}
\begin{document}
\begin{center}{\Large{\bf Standing spin wave mode in RFIM at T=0:}\\ 
{\bf Patterns and athermal nonequilibrium phases}}\end{center}

\vskip 1cm

\begin{center}{\it Muktish Acharyya}\\
{\it Department of Physics, Presidency University,}\\
{\it 86/1 College street, Calcutta-700073, INDIA.}\\
{E-mail:muktish.physics@presiuniv.ac.in}\end{center}

\vskip 2cm

\noindent {\bf Abstract:} The dynamical responses of random field Ising model
at zero temperature, driven by standing
magnetic field wave, is studied by Monte Carlo simulation
in two dimensions. The three
different kinds of distribution of quenched random field are used here,
uniform, bimodal and Gaussian. In all cases, three distinct dynamical
phases were observed, namely, the {\it pinned}, {\it structured} and
{\it random}. In the pinned phase no spin flip is observed. In the structured
phase standing spin wave modes are observed. The random phase is shown with no
observed regular pattern. For a fixed value of the 
amplitude of the standing magnetic 
field wave, in the region of small quenched field, the system remains in 
a pinned phase. In the intermediate range of values of random field, a
standing spin wave mode (structured phase) is observed. The regular pattern
of this spin wave mode disappears for higher values of random field yielding
a random phase. The comprehensive phase baundaries are drawn in all three
cases. The boundary of pinned phase are analytically calculated for
uniform and bimodal types of quenched random fields.

\vskip 1cm


\newpage

\noindent {\bf 1. Introduction:}

The random field Ising model (RFIM) is a simple model to understand various physical phenomena,
like, hysteresis\cite{book}, Barkhausen noise
\cite{sethna1} observed in ferromagnetic materials, avalanche\cite{perkovic},
return point memory effects etc. The RFIM
was studied theoretically with intense attention. RFIM was solved exactly\cite{dhar}
on Bethe lattice and the qualitative behaviour of magnetisation as a function of 
external magnetic field was found to depend on the coordination number of Bethe lattice.
The effect of coordination number on the nonequilibrium critical point in RFIM was
studied\cite{shukla1} in details. Hysteresis in RFIM with asymmetric distribution of quenched
random fields in the limit of low disorder was studied\cite{sanjib1} and the spin flip was
found to be related to bootstrap percolation. Athermal hysteresis was also studied\cite{lobisor}
in antiferromagnetic RFIM recently. The statistics\cite{tadic1} and the dynamical critical
behaviour of avalanches\cite{tadic2} was studied in RFIM. All these studies mentioned here
are static in a sense that the time dependence of any quantity is studied. 

The dynamical behaviours of the driven RFIM are also quite interesting.
The RFIM has also served to study the dynamics of domain wall. The motion of domain wall
by magnetic field in a 2D ultrathin Pt/Co/Pt film (showing perpendicular anisotropy and
quenched disorder which is analogous to RFIM) was studied\cite{lemerele} experimentally
by magneto optical polar Kerr imagning technique. The motion of moving interface
was analyzed numerically\cite{roters1} in the RFIM driven by external magnetic field.
Later, the depinning transition of driven interface was studied\cite{lubeck}. The creep
motion of interface in driven RFIM was studied
\cite{dong} and the nonlinear field-velocity relationship
was found.

All the studies mentioned above have a very common feature. Firstly, the external driving
magnetic field was varied is a quasi-static manner. Secondly, the driving magnetic field
is uniform over the lattice at any given instant of time. The RFIM also shows interesting
dynamical behaviours if the uniform driving field varies rapidly. The dynamic hysteresis
and dynamical athermal phase transition was already studied\cite{ma1} in RFIM. More
interesting news comes from the fact when RFIM is driven by a field having both spatio-temporal
variation. Very recently, the RFIM was studied\cite{ma2}
 in the presence of propagating magnetic field
wave. Various nonequilibrium phases are observed\cite{ma2} and a phase boundary was drawn.
In this paper, the dynamical responses of RFIM driven by standing magnetic field wave
is studied in details. The paper is organised as follows: the model and simulation are
discussed in section-2, numerical results are reported in section-3 and the paper ends with
a summary of the work in section-4. 

\vskip 2cm

\noindent {\bf 2. Model and simulation:}

The Ising model on a square lattice ($L\times L$), 
in the presence of quenched random magnetic field and driven by
a magnetic field having spatio-temporal variation, can be described 
by the Hamiltonian,
\begin{equation}
H = -J\Sigma s(x,y,t)s(x',y',t) -\Sigma h_r(x,y)s(x,y,t) -\Sigma
h_s(x,y,t)s(x,y,t).
\end{equation}

\noindent Where $s(x,y,t)=\pm1$ is Ising spin at site (x,y) and at time $t$.
$J>0$ is ferromagnetic spin-spin interaction (nearest neighbour only) strength,
$h_r(x,y)$ is quenched random field and $h_s(x,y,t)$ is external magnetic
field having spatio-temporal variation. In the present study, $h_s(x,y,t)$
represent the value of the external magnetic field at lattice site (x,y) and
at time $t$. The standing magnetic wave field $h_s(x,y,t)$ can be represented
as

\begin{equation}
h_s(x,y,t) = h_0 {\rm Sin}(2\pi f t){\rm Sin}({{n\pi y} \over {L}}).
\end{equation}

\noindent Where, $h_0$ and $f$ represent the amplitude and frequency
of the standing 
magnetic field wave respectively. Here, $n$ represents the number of antinodes
in the standing magnetic wave. The 
standing magnetic field wave spreads along the y direction only. 
The preiodic boundary conditions are applied in both directions.

The distributions of the quenched
random field $h_r(x,y)$, are considered here of folllowing three types:

(a) The uniformly distributed between $-w$ and $+w$, with prbability
$P_u(h_r) = {{1} \over {2w}}$ if $-w \leq h_r \leq +w$ and 0 otherwise.

(b) Having a bimodal probability distribution
\begin{equation}
P_b(h_r)=0.5\delta(h_r-w) + 0.5\delta(h_r+w).
\end{equation}
\noindent where $\delta$ reprensents the Dirac delta function.

(c) The normally distributed with 0 mean and standard deviation $\sigma$
(=$2w$ here) with probability
\begin{equation}
P_n(h_r) = {{1} \over {\sqrt{2\pi \sigma^2}}} e^{-{{{h_r}^2} \over {2\sigma^2}}}
\end{equation}

This RFIM driven by standing magnetic field wave is studied by Monte
Carlo simulation at zero temperature. The zero temperature Metropolis
dynamics is cosidered here in the manner that the spin only flips if
it lowers the energy\cite{glauber}. Starting from an initial configuration as
\begin{equation}
s(x,y,t=0)=+1 ~~~ \forall ~~~x,y.
\end{equation}
\noindent The parallel updating rule is used here. $L^2$ such updates of
spins constitutes the unit time step, i.e. Monte Carlo Step (MCS). Throughout
the simulation the frequency of the standing magnetic field wave is kept
constant ($f=0.01$). So, time period $\tau=100$
 MCS is required to have a complete 
temporal cycle of
the standing magnetic field wave. For n loop standing wave, one would get
n number of antinodes in the length $L$. As a result the wavelength of the
standing magnetic field wave $\lambda=2L/n$. Here, the simulation is done
and the results are shown for $n=4$ and $n=2$. The linear size of the square 
lattice is chosen $L=100$ here. The wavevector, $k={{2\pi} \over {\lambda}}$.

\vskip 2cm

\noindent {\bf 3. Numerical Results:}

Here, the dynamical responses of the model are studied for three different
kinds of the distributions of quenched random field. For uniformly distributed
(eqn-3) quenched random field, depending on the values of $w$, $h_0$ mainly three
different kinds of dynamical or nonequilibrium phases are observed.  
For $h_0=1.0$ and $w=1.0$, the RFIM remains pinned, i.e., no spin flips was
observed. This is called {\it pinned} phase. For the higher values of
$h_0$ and $w$, namely $h_0=2.0$ and $w=3.0$, a {\it structured} phase was
observed. Here, the band like sturcture of the clusters of the spins 
are formed. These bands oscillates periodically as the fields changes sign
in the standing magnetic field wave. 
A typical phase of this kind may be found in a video presentation in
http://youtu.be/5pzW3chzYnw. Note that the value of the $n$ is taken here
equals to 4. That corresponds to the presence of 4 antinodes in the standing
magnetic field wave.
For $h_0=1.0$ and $w=5.0$, a random
spin configuration (with no specific structure) is found. This is called
{\it random} phase. Figure-1 shows snapshots of such dynamical phases. 

For a particular value of $h_0$, the three phases (mentioned above) 
are found by varying the strength of the quenched random field $w$.
The system remains in a pinned phase untill a value of $w$ is reached.
After that, it remains in the structured phase and ultimately it reaches
the random phase for much higher values of $w$.

All these nonequilibrium phases can be characterise by the following
quatitites: the dynamic structure factor 
$S(k,t)=(1/L) \int_0^L m(y,t) e^{iky} dy$, where the line magnetisation along
y direction is defined as $m(y,t) = (1/L) \int_0^L s(x,y,t) dx$. Here, the
integrals are essentially the summation over discrete lattice point. The
time average magnetisation over the full cycle of the 
standing magnetic field wave is defined as $Q={{1} \over {L \tau}}
\oint m(y,t) dt$. Here also, the integral is essentially the discrete 
sum over the MCS. All these quantities are obtained by averaging over
100 different samples of quenched random field.

In the pinned phase, the structure factor $S(k,t)$ will be zero and $Q$
will be unity. However, in the structured phase $S(k,t)$ is nonzero and
$Q$ will be less that unity (due to the presence of spins of opposite
sign). In the random phase, where no specific pattern is present and 
approximately 50 per cent of the total number of spins have opposite sign.
As a result, $Q$ and $S(k,t)$ will be nearly equal to zero. For a fixed
value of $h_0$ the transition from pinned to random via the structured 
phase was observed by studying the $S(k,t)$ and $Q$ as functions of $w$.
In the case of uniformly distributed quenched random field, Figure-2 
shows such variations for $h_0=2.0$ and $h_0=3.0$. Here, the $n$ is chosen
equal to 4. For small values of $w$, the system 
remains in pinned phase ($|S(k,y)|=0.0$ and $Q=1.0$ in fig.2(a) and fig2.(b)).
Above a particular value of $w$, the system transits to the sructured (
middle of Fig-1) phase, where $|S(k,t)|$ shows a large nonzero value
and $Q$ becomes less than unity. 
This transition point can be marked by a sharp maximum and minimum of the
derivatives ${{d(|S(k,t)|)} \over {dw}}$ and ${{dQ} \over {dw}}$
respectively (fig.2(c) and fig.2(d)). This transition point is $w_p$.
As the $w$ increases, both $Q$ and $|S(k,t)|$
decreases and eventually vanishes in the random phase. It is observed that
structured phase persists as long as $|S(k,t)|$ remains larger than 0.1. The 
structure completely disappears for $|S(k,t)| < 0.1$ and the system gets
into random phase. The transition from structured to random phases are marked
by $|S(k,t)| < 0.1$. This transition point is called $w_s$. 
It is noted that, the transition values of the queched random field, 
$w_p$ and $w_s$ depend on the value of $h_0$. The $w_p$ decreases as $h_0$
increases. Physically, larger amount of standing wave field is required to
flip the spin for weak quenched random field. On the other hand, $w_s$ 
increases as $w$ increases upto a certain value of $w$. 

The comprehensive phase boundaries (pinned-structured and stuctured-random)
can be plotted in the plane described by $h_0$ and $w$. It is shown in 
fig-3. It may be noted the the bounadry of pinned and structured phases
is linear. This can be realised as follows: the spin flip will not be possible
until the local field 
is negative. The local field $F=4-(w+h_0)$ remains positive as long as
$w+h_0 < 4$, preventing the spin flip. So, the boundary of pinned phase
is nothing but a straight line $w+h_0=4$. The phase boundaries are also obtained
for $n=2$. No considerable difference was observed.

The similar studies were done for bimodal distribution (eqn-4) of quenched random 
field. The qualitative behaiours are similar to that observed in the case
of uniformly distributed quenched random field. 
Figure-4 shows the plots
of $|S(k,t)|$, $Q$ 
and the derivatives ${{d(|S(k,t)|)} \over {dw}}$ and ${{dQ} \over {dw}}$
as functions of $w$ for two different values of $h_0$. 
The comprehensive
phase baundaries (for $n=4$ and $n=2$) are
 shown in figure-5. Here, the boundary of pinned-structured
phase is linear and same argument may be applied to analyse it. However,
the boundary of structured-random phase for higher values of $h_0$ is a 
line $w=4.0$. As the distribution is bimodal, $w$ can be +4 or
-4. So, for larger values of $h_0$, $w=4.0$ will randomise the spin structure.

The normally distributed (eqn-5) quenched random field was also used to study the
dynamical or nonequilibrium behavious of RFIM at $T=0$. Here also, 
qualitatively similar kind of behaviours were observed. 
Figure-6 shows the plots
of $|S(k,t)|$, $Q$ 
and the derivatives ${{d(|S(k,t)|)} \over {dw}}$ and ${{dQ} \over {dw}}$
as functions of $w$ for two different values of $h_0$. 
However, the region
of pinned phase is much smaller than that for other two (uniform and
bimodal) cases. The quality of the data is not quite conclusive to get any
analytic form of the phase boundary. The comprehensive phase boundaries
(for $n=4$ and $n=2$) are shown in figure-7.

\vskip 2cm

\noindent {\bf 4. Summary:}

The nonequilibrium behaviours of the random field Ising ferromagnet, at zero temperature, 
driven by standing magnetic field wave are studied by Monte Carlo simulation in two dimensions.
The three different kinds, namely, uniform, bimodal and Gaussain, of distributions of
the quenched random fields are considered here. The random field Ising ferromagnet is driven
by an additional standing magnetic field wave of well defined amplitude, frequency and
wavelength. The wavelength depends on the number of antinodes are present in the standing
wave. In the present study, only 2-loop and 4-loop standing waves are considered.
Starting from an initial spin configuration having all spins are up, the zero temperature
Metropolis dynamics are employed.
In this case, the spin only flips if its lowers the energy. The rule parallel updating is used here. 

In the dynamical steady state, the system shows three distinct nonequilibrium phases depending
on the values of the amplitude of standing magnetic field wave and the strength of quenched
random field. For a fixed value of the amplitude of the standing magnetic field wave, in the
limit of weak quenched random field, the system remains in a pinned phase. In this phase,
all spins remain up as considered in the initial configuration. In the intermediate values of
the quenched random field, a standing spin wave mode is observed. Here, the bands of up and
down spins formed alternately and oscillates. This is different from the propagation mode
studied\cite{ma2} earlier in RFIM driven by propagating magnetic field wave. Where the bands
of alternate up and down spins moved in the direction of the propagating magnetic field wave.

The comprehensive phase boundaries are drawn in the plane formed by the strength of random field
and the amplitude of standing magnetic field wave. These phase boundaries are qualitatively
similar in three different kinds of distributions of quenched random field. The boundaries
of pinned phases in the cases of uniform and bimodal distribution can be obtained analytically.

The compound ${\rm Li} {\rm Ho_x} {\rm Y_{1-x}} {\rm F_4}$ can be modelled by random field Ising
model. On applying standing magnetic field wave, a possible way to see these effects experimentally, 
may be the relaxation experiments using SQUID magnetometer \cite{ltp}.

\vskip 2cm
\noindent {\bf Acknowledgements:} 
Author would also like to thank Ritaban Chatterjee for his help to prepare
the video of standing spin wave modes.

\newpage
\setlength{\unitlength}{0.240900pt}
\ifx\plotpoint\undefined\newsavebox{\plotpoint}\fi
\sbox{\plotpoint}{\rule[-0.200pt]{0.400pt}{0.400pt}}%


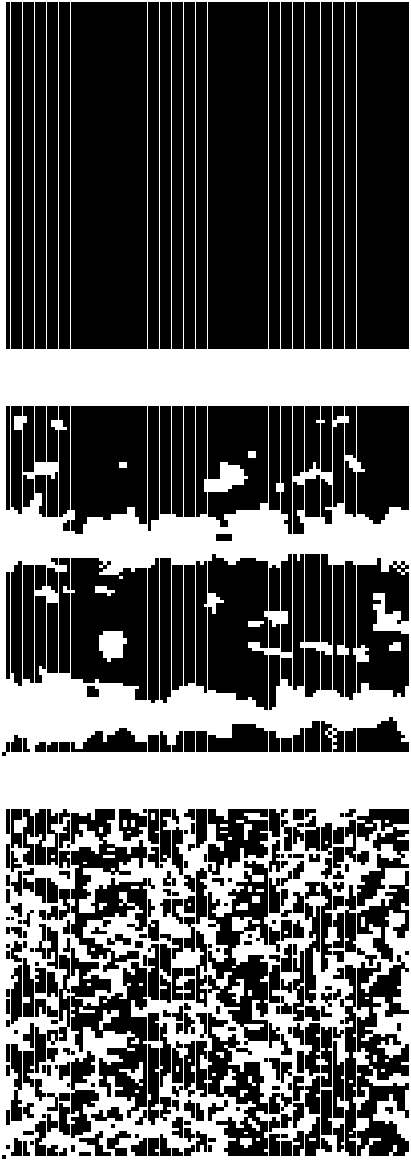
\noindent {\bf Fig-1.} The snapshots (taken after $t=100$ MCS)
of spin configuration for uniform distribution 
of quenched random field. The black dots represent $s(x,y,t)=+1$.
 The pinned phase is shown (for $h_0=1.0$,
$w=1.0$) in the top. The structured (for $h_0=2.0$, $w=3.0$) 
phase is shown in the middle. The bottom one shows the random (for $h_0=1.0$,
$w=5.0$) phase. Here, $L=100$ and 4-loop standing magnetic field wave is
considered in each case. The dynamics of the structured phase can be found 
in http://youtu.be/5pzW3chzYnw.
 
\newpage
\setlength{\unitlength}{0.240900pt}
\ifx\plotpoint\undefined\newsavebox{\plotpoint}\fi
\sbox{\plotpoint}{\rule[-0.200pt]{0.400pt}{0.400pt}}%


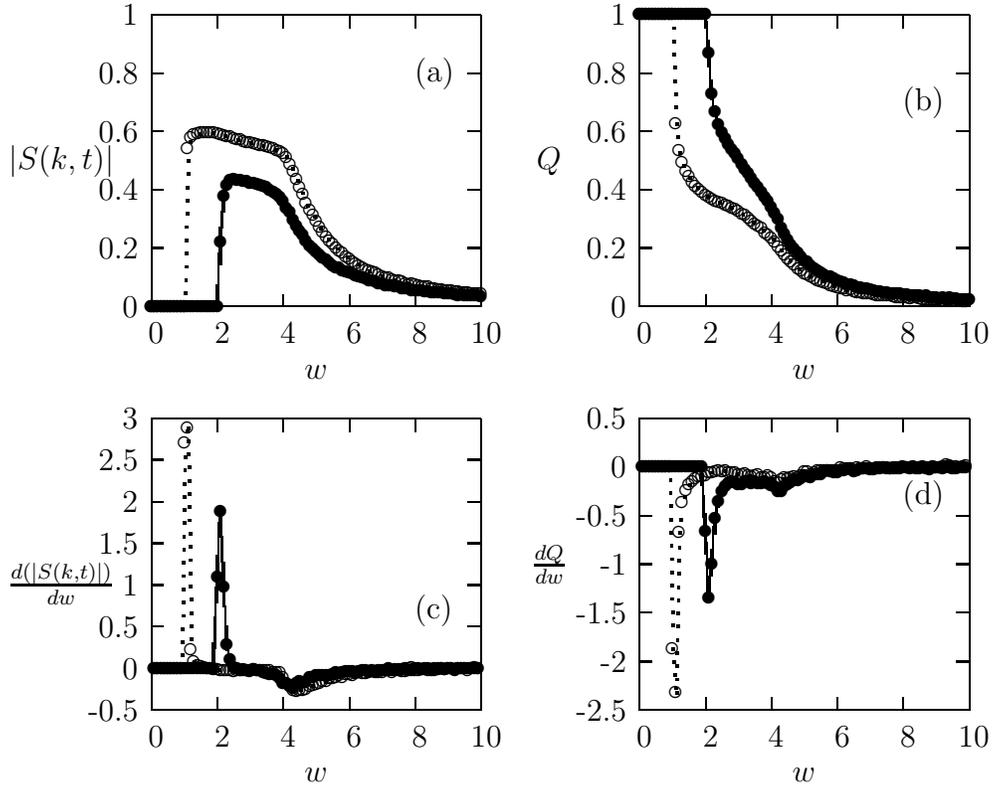
\noindent {\bf Fig-2.} The $|S(k,t)|$, $Q$, ${{d(|S(k,t)|)} \over {dw}}$ and ${{dQ} 
\over {dw}}$ are plotted against $w$ for uniform distribution of quenched
random field for two different values of $h_0$ and 4-loop of standing
magnetic field wave. Different symbols represent different values of $h_0$.
$(\bullet)$ represents $h_0=2.0$ and $(\circ)$ represents $h_0=3.0$. 
Solid and dotted lines are just connecting the data points.

\newpage
\setlength{\unitlength}{0.240900pt}
\ifx\plotpoint\undefined\newsavebox{\plotpoint}\fi
\sbox{\plotpoint}{\rule[-0.200pt]{0.400pt}{0.400pt}}%


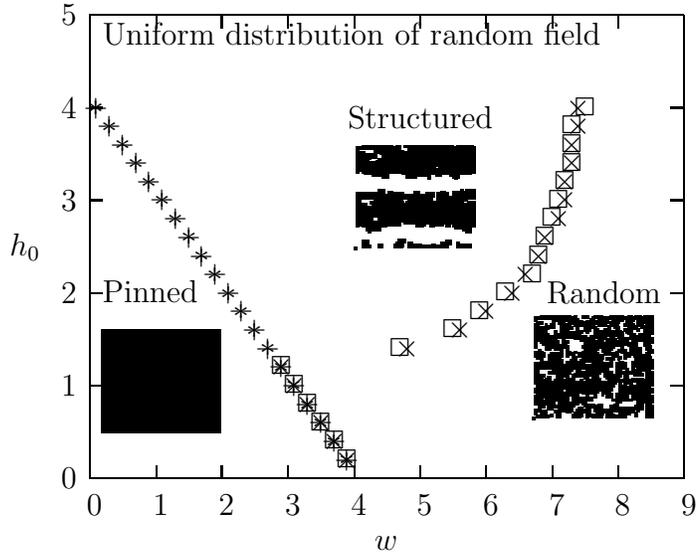
\noindent {\bf Fig-3.} The phase diagram for uniform distribution of quenched random field.
The boundaries of pinned-structured ($+$) phase and the  
structured-random ($\times$)
phase are shown for $n=2$. The boundaries of pinned-structured ($\ast$)
phases and the structured-random ($\Box$) 
phases are also shown for $n = 4$. The typical spin configurations 
(for 60$\times60$ lattice) are show in the insets for three different phases.
The pinned phase ($h_0=1.0$, $w=1.0$), the structured phase ($h_0=2.0$, $w=3.0$)
and the random phase ($h_0=1.0$, $w=5.0$) are shown for $n=4$.

\newpage
\setlength{\unitlength}{0.240900pt}
\ifx\plotpoint\undefined\newsavebox{\plotpoint}\fi
\sbox{\plotpoint}{\rule[-0.200pt]{0.400pt}{0.400pt}}%


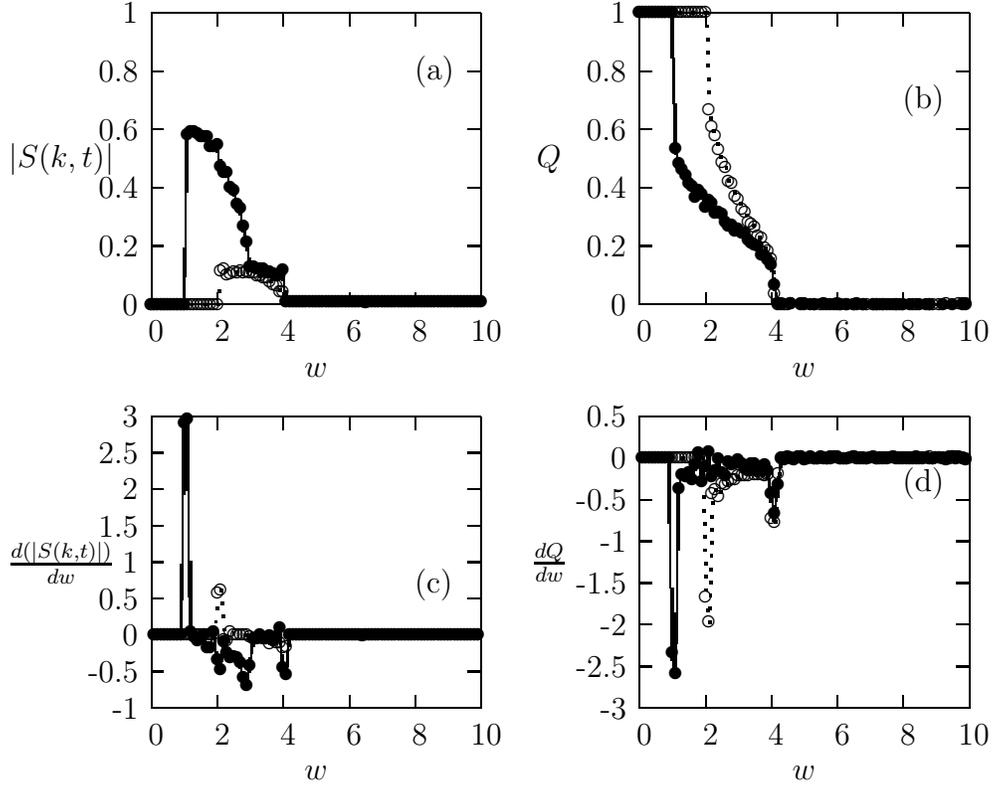
\noindent {\bf Fig-4.} The $|S(k,t)|$, $Q$, ${{d(|S(k,t)|)} \over {dw}}$ and ${{dQ} 
\over {dw}}$ are plotted against $w$ for bimodal distribution of quenched
random field for two different values of $h_0$ and 4-loop of standing
magnetic field wave. Different symbols represent different values of $h_0$.
$(\bullet)$ represents $h_0=3.0$ and $(\circ)$ represents $h_0=2.0$. 
Solid and dotted lines are just connecting the data points.

\newpage
\setlength{\unitlength}{0.240900pt}
\ifx\plotpoint\undefined\newsavebox{\plotpoint}\fi
\sbox{\plotpoint}{\rule[-0.200pt]{0.400pt}{0.400pt}}%


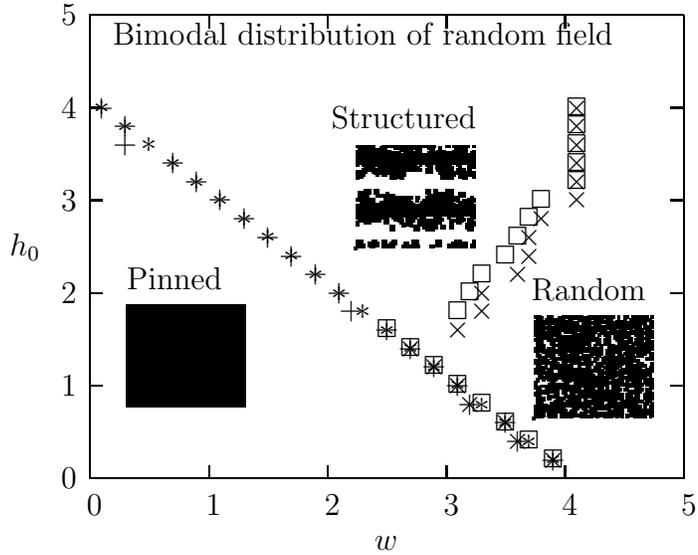
\noindent {\bf Fig-5.} The phase diagram for bimodal distribution of quenched random field.
The boundaries of pinned-structured ($+$) phase and the  
structured-random ($\times$)
phase are shown for $n=2$. The boundaries of pinned-structured ($\ast$)
phases and the structured-random ($\Box$) 
phases are also shown for $n = 4$. The typical spin configurations 
(for 60$\times60$ lattice) are show in the insets for three different phases.
The pinned phase ($h_0=1.0$, $w=1.0$), the structured phase ($h_0=2.4$, $w=2.0$)
and the random phase ($h_0=1.0$, $w=4.0$) are shown for $n=4$.

\newpage
\setlength{\unitlength}{0.240900pt}
\ifx\plotpoint\undefined\newsavebox{\plotpoint}\fi
\sbox{\plotpoint}{\rule[-0.200pt]{0.400pt}{0.400pt}}%


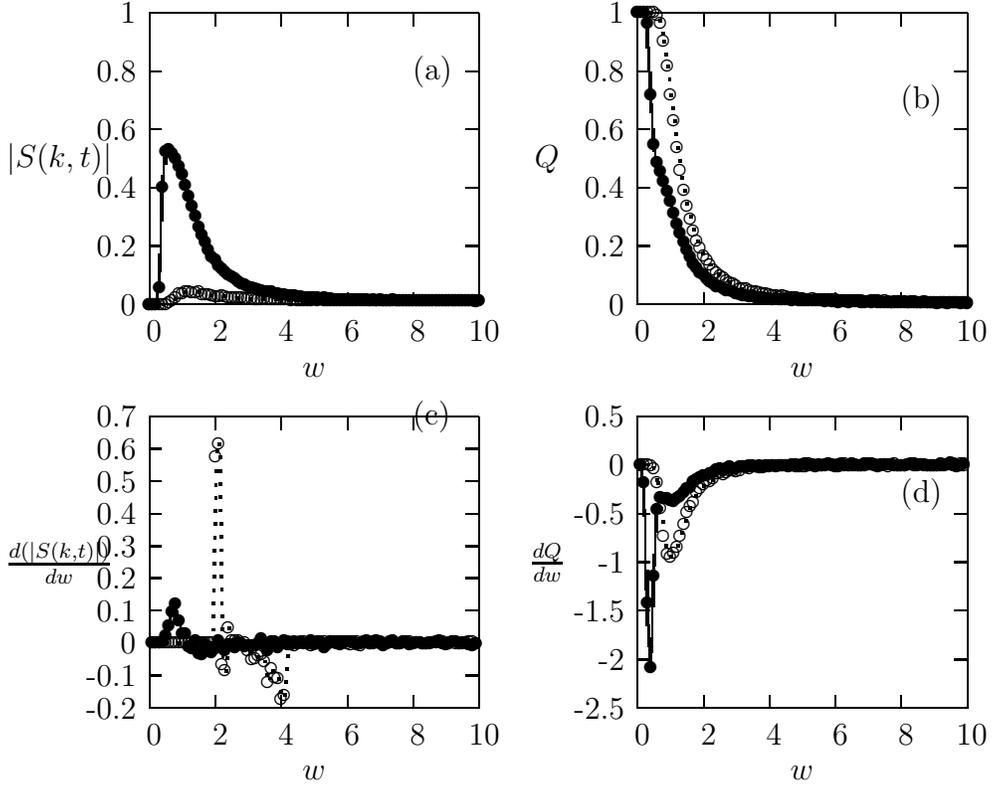
\noindent {\bf Fig-6.} The $|S(k,t)|$, $Q$, ${{d(|S(k,t)|)} \over {dw}}$ and ${{dQ} 
\over {dw}}$ are plotted against $w$ for Gaussian distribution of quenched
random field for two different values of $h_0$ and 4-loop of standing
magnetic field wave. Different symbols represent different values of $h_0$.
$(\bullet)$ represents $h_0=2.2$ and $(\circ)$ represents $h_0=1.0$. 
Solid and dotted lines are just connecting the data points.

\newpage
\setlength{\unitlength}{0.240900pt}
\ifx\plotpoint\undefined\newsavebox{\plotpoint}\fi
\sbox{\plotpoint}{\rule[-0.200pt]{0.400pt}{0.400pt}}%


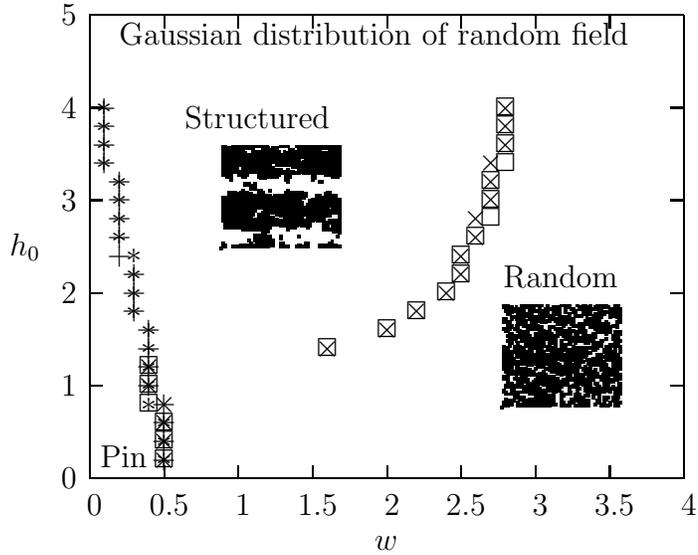
\noindent {\bf Fig-7.} The phase diagram for Gaussian distribution of quenched random field.
The boundaries of pinned-structured ($+$) phase and the  
structured-random ($\times$)
phase are shown for $n=2$. The boundaries of pinned-structured ($\ast$)
phases and the structured-random ($\Box$) 
phases are also shown for $n = 4$. The typical spin configurations 
(for 60$\times60$ lattice) are show in the insets for two different phases.
The structured phase ($h_0=2.0$, $w=1.0$)
and the random phase ($h_0=1.0$, $w=3.0$) are shown for $n=4$. The 
configuration of pinned phase is similar to that for other two cases (uniform
and bimodal) and could not be shown here due to the limitation of 
the region of space in
pinned phase.
\end{document}